\documentclass[12pt]{iopart}
\usepackage[latin9]{inputenc}
\usepackage{color}
\usepackage{verbatim}
\usepackage{amssymb}
\usepackage{graphicx}
\usepackage{hyperref}

\begin{document}

\title{Holographic optical traps for atom-based topological Kondo devices}

\author{F~Buccheri$^1$, G~D~Bruce$^2$, A~Trombettoni$^{3,4}$, D~Cassettari$^2$, H~Babujian$^{1,5}$, V~E~Korepin$^{1,6}$ and P~Sodano$^{1,7}$}

\address{$^1$ International Institute of Physics, Universidade Federal 
do Rio Grande do Norte, 59078-400 Natal-RN, Brazil}

\address{$^2$ SUPA School of Physics and Astronomy, 
University of St Andrews, North Haugh, St Andrews, KY16 9SS, UK}

\address{$^3$ CNR-IOM DEMOCRITOS Simulation Center, Via Bonomea 265, I-34136
Trieste, Italy}
\address{$^4$ SISSA and INFN, Sezione di Trieste, Via Bonomea 265, I-34136 
Trieste, Italy}

\address{$^5$ Yerevan Physics Institute, Alikhanian Brothers 2,
Yerevan, 375036, Armenia}

\address{$^6$ C. N. Yang Institute for Theoretical Physics, 
Stony Brook University, NY 11794, USA}

\address{$^7$ Departemento de F\'isica Teorica e Experimental,
Universidade Federal do Rio Grande do Norte, 59072-970 Natal-RN, Brazil}

\begin{abstract}
The topological Kondo (TK) model   
has been proposed in solid-state quantum devices 
as a way to realize non-Fermi liquid behaviors in a controllable setting.
Another motivation behind the TK model proposal is the demand  
to demonstrate the quantum dynamical properties of 
Majorana fermions, which are at the heart of their potential use in 
topological quantum computation.
Here we consider a junction of crossed Tonks-Girardeau gases arranged
in a star-geometry (forming a $Y$-junction), and we 
perform a theoretical analysis of this system showing that 
it provides a physical realization of the topological Kondo model in the realm 
of cold atom systems. 
Using computer-generated holography, we experimentally implement
a $Y$-junction suitable for atom trapping, with controllable and independent parameters. 
The junction and the transverse size of the atom waveguides
are of the order of $5$ $\mu$m, leading to favorable estimates for 
the Kondo temperature and for the coupling across the junction. 
Since our results show that all the required  
theoretical and experimental ingredients are available, this provides 
the demonstration of an ultracold atom device that may 
in principle exhibit the topological Kondo effect.
\end{abstract}
%{\it Keywords\/}:

%\maketitle

\section{Introduction}

The topological Kondo (TK) model is a device exploiting the degeneracy
of the spectrum of a system in which a set of Majorana modes is present
\cite{Leijnse,Beenakker,Alicea,Stanescu}.
It  has been recently introduced 
with the goal of realizing the physics of multichannel Kondo models in 
solid state devices \cite{BeriCooper2012}. 
Moreover, a major motivation in the quest of TK devices 
is the realization of hardware devices
for quantum computation \cite{Preskillnotes,Nielsenchuang}. The rationale 
is that the manipulation of anyonic excitations is at the heart 
of topological quantum computation \cite{NayakTQC}.
Hence, proper control of a device with Majorana modes gives the possibility 
of performing a wide class of quantum information tasks \cite{NayakTQC,Hyart,vanHeck,Sau2011,Milestones}.
In this respect, the TK device can be considered as the elementary unit 
of a variety of hardware setups. 
In general, if the states of Majorana zero modes can be measured in
well-controlled experiments, this could pave the way
towards the realization of a topological
quantum computer \cite{DasSarma}, together with the possibility of
manipulating the Majorana modes themselves, or performing measurement
of certain observables involving the Majorana modes on special geometries.
It appears that the control of the TK unit 
may provide an important resource 
to demonstrate the properties of Majorana modes and in turn 
control their dynamics, which is instrumental in their forseen use in 
topological quantum computation \cite{NayakTQC}.
The possibility of such a control is currently one of the main goals 
of different lines of research in physics.

In this paper, we propose a new realization of the TK model in an 
highly controllable setup using cold atoms. We also present 
an experimental implementation of a suitably engineered holographic optical
trap, which constitutes the central element of our realization.

The TK effect can be obtained, at low temperature, in a setup where a set of localized Majorana modes is hosted on a central island. One end of the (effectively one-dimensional) external wires is then connected to the central island,
in such a way that the tunneling of electrons to or from the island
can change the state of the set of Majorana degrees of freedom 
\cite{Fu2010}.
The Majorana modes can then be seen as nonlocally encoding a degree of freedom, which
is much alike that of a localized impurity interacting with a gas of fermions. In this solid state proposal, 
the actual realization of TK devices would 
provide a playground for the investigation 
of the properties of non-Fermi liquid critical points.
The study of non-Fermi liquids plays a prominent role in the theory of strongly-correlated
solid-state systems \cite{hewson1997kondo}.

To date, experimental realization of solid state TK devices has remained elusive. 
The reasons, in solid state proposals, are many.
Among these, the fact that the single-particle tunneling onto the central 
island at temperatures 
which are not negligible compared to its charging energy may spoil the TK effect
by mixing parity sectors \cite{Buccheri2015}. Moreover, in realistic devices, 
the lifetime of localized Majorana modes is necessarily finite
\cite{Higginbotham2015,Rainis2012}. 
The quasiparticle poisoning, connected to the presence of 
an unpaired electron within the superconducting substrate, is the phenomenon which most contributes 
to shorten the lifetime of localized Majorana modes.
A considerable effort is in progress in this direction \cite{ZazuPois}. %Pluggesurface,

Complementary to such efforts, alternative approaches 
are highly desirable, in particular if an accurate control of 
the parameters characterizing the wires and their coupling with the
Majorana modes can be achieved. 
For this reason, in this paper, we not only propose how to realize the TK Hamiltonian with cold atoms 
in a {\em Y-junction}, but we also show that it is possible to experimentally implement the $Y$-junction 
needed for this purpose.
This junction may provide a basic component for connecting atom guides 
and more generally for atomtronics \cite{Ryu2015}.

Cold atom setups emerge as a natural candidate 
to simulate low-energy Hamiltonians \cite{Thomas}.
The architecture needed to implement the TK Hamiltonian 
is that of a suitably engineered quantum system 
trapped in a set of effectively one-dimensional waveguides,
joined together in a central region. 
Such a geometry is often referred to as a $Y$-junction (or also $T$-junction). 
It is possible to map \cite{BeriCooper2012} the effective TK Hamiltonian to 
$Y$-junction systems \cite{ChamonPhysRevLett.91.206403,Chamon}. On the 
other hand $Y$-junctions of Ising chains are, in turn, a physical realization of the 
TK Hamiltonian \cite{Tsvelik:2014Ising}.
In a cold atom realization, it would then be possible to tune the parameters of quantum one-dimensional systems 
merging in the junction, as well as the shape and characteristics of the junction, through 
the use of controllable traps.
Moreover, in the proposed experimental setup,
the Majorana modes will be nonlocally encoded in some bosonic bulk 
fields and, therefore, will not be affected by any of the above-mentioned drawbacks
characteristic of solid state devices.

Two difficulties have to be 
overcome in order to have a TK device in the realm of cold atoms: 
{\em i)} one should find a low-dimensional cold atom system having 
an effective low-energy Hamiltonian which matches that of the TK model; {\em ii)} 
one should have a reliable and controllable realization of the system 
geometry. It is clear that the ``wires'' of solid state proposals correspond 
to ``waveguides'' in cold atom setups in which the atoms are tightly confined: 
in the following, we will use both terms (``wires'' and ``waveguides'') to 
denote the different branches of the $Y$. 

Among the different lines of research in the field of quantum simulations 
with cold gases, a very active one and relevant for our purposes  
is provided by the study of one-dimensional gases. 
There are several reasons for such an interest: for low dimensional systems 
in the last few decades powerful tools have been developed, ranging 
from bosonization to integrability methods like the Bethe ansatz 
\cite{korepin1997quantum,Boso}. Furthermore these experimental configurations 
are ideal to test different approaches
developed for systems where integrability is broken 
and to study controllable deviations from 
integrable models, e.g. achieved by connecting together several
one-dimensional systems. 
One way of coupling one-dimensional systems is to glue them in a point or, 
more realistically, in a small part of them, creating a $Y$-like 
geometry, which is an essential ingredient to realize 
the TK effect with cold atoms. 
This way 
of coupling one-dimensional chains is possible in some solid state systems, e.g. 
carbon nanotubes \cite{nanotube}. 

The implementation of such a layout with cold 
atoms would open stimulating possibilities in the light of 
providing the $Y$-shape of the junction needed 
in the TK effect. 
Recently, part of a bosonic condensate in a quasi-one-dimensional
optical trap has been split into two branches \cite{Ryu2015}, 
generating the nontrivial geometry of an $Y$-junction.
However, despite this and other considerable progress in the field, 
to date no experimental realization of a stable and controllable 
$Y$-geometry is known. In particular, it is required 
that several waveguides merge in a well defined region, that 
the tunnelings of atoms among different waveguides be tunable and that 
the low-energy Hamiltonian of the atoms in each 
waveguide be effectively one-dimensional.

One method to realize unusual optical trapping geometries is computer generated holography, in which a phase modulation is applied to the trapping light such that a desired intensity distribution is realized in the far field. Holographic optical traps provide a flexible tool to tailor the 
potential experienced by neutral atoms, and have been employed in %demonstrated 
experiments ranging from single atoms 
\cite{Bergamini04,XHe09,Nogrette14} to Bose-Einstein 
condensates \cite{Boyer06,Zoran2013}. The development of 
numerical \cite{Pasienski08,Harte14} and experimental 
\cite{Bruce11,Gaunt12,Bruce15,Bowman15} techniques which 
allow the creation of smooth light profiles has given 
significant freedom in engineering different trapping geometries, 
including proposals in investigations of superfluidity 
\cite{Bruce11}, atomtronics \cite{Gaunt12} and entropy-engineering 
of ultracold Fermi gases \cite{Bowman15}. 

For the implementation of the TK model in the framework of atomtronics,
 two preliminary obstacles have to be tackled.
The first is the realization of the junction itself, with
fabrication parameters that allow functionality and tunability.
The second is the understanding of the low-temperature properties of this junction,
when bosonic atoms are loaded in the trap.
In this paper, we provide a path to overcome both these obstacles
and we show that holographic optical traps 
can be exploited for a systematic implementation of the junction geometry for TK devices.
Our results are the following:
\begin{itemize}
\item We show that 
when the interacting Bose gases in the wires 
are in the Tonks-Girardeau (TG) limit, the $Y$-junction 
provides a physical realization of the TK Hamiltonian 
\cite{BeriCooper2012,Altland2014multi,Tsvelik:2014Ising,Altland2014,Buccheri2015}. 
This identification is fruitful for two reasons. 
Firstly, it opens the possibility 
of studying the TK model and excitations 
at the junction with cold bosons in an highly controllable experimental 
setup. Secondly, recent results obtained from the theory 
of the TK model allow to write 
exact expressions for thermodynamical quantities, as well as
to characterize the transport across the central region. 
In all these results the formalism associated with Majorana modes 
plays a central role (see also the discussion in section \ref{sec:conclusions})
and we will describe how to detect the experimental signatures of the
Majorana fermion physics in the system under study.
\item  We experimentally demonstrate the possibility 
of creating $Y$-junctions for cold bosons using 
holographic techniques. We show that one can realize junction widths of 
the order of 5 $\mu$m, with tunable 
hopping parameters, using blue- and red-detuned laser potentials (respectively 
providing repulsive and attractive potentials on the atoms 
\cite{Grimm_00}).
Given that the one-dimensional regime for cold bosons is reachable \cite{Yurovsky08,Bouchoule09,Cazalilla11}
and that TG gases have been widely studied in experiments \cite{Paredes2009,Kinoshita2004}, 
the realization of holography-based $Y$-junctions provides a 
proof-of-principle of an 
atom-based holographic device that potentially may exhibit
the TK effect.
\end{itemize}

\section{The topological Kondo model}
\label{sec:TKreview}
The Hamiltonian of the topological Kondo model (or Coulomb-Majorana box) is:
\begin{eqnarray}\label{eq:TopologicalKondoHamiltonian}
 H&=&-i\frac{\hbar v_{F}}{2\pi}\sum_{\alpha=1}^{M} \int dx \psi_{\alpha}^{\dagger}(x)\partial_{x}\psi_{\alpha}(x)
 + \sum_{\alpha\ne\beta}\lambda_{\alpha\beta}\gamma_{\alpha}\gamma_{\beta}\psi_{\beta }^{\dagger}(0)\psi_{\alpha}(0)
 \;.
\end{eqnarray}
Here $\psi_{\alpha}(x)$ are the (complex) Fermi fields describing electrons in the wires $\alpha=1,\ldots,M$
and satisfying canonical anticommutation relations
\begin{equation}
 \left\lbrace\psi_{\alpha}(x),\psi_{\beta}(y)\right\rbrace = 0 \qquad\qquad \left\lbrace\psi_{\alpha}(x),\psi_{\beta}^\dagger (y)\right\rbrace = \delta_{\alpha,\beta}\delta(x-y)
 \;,
\end{equation}
while the $\gamma_{\alpha}=\gamma_{\alpha}^\dagger$ are Majorana fields constrained in a box connected with the wires and satisfying the Clifford algebra
 \begin{equation}\label{eq:PoissonAlgebra}
  \left\lbrace\gamma_{\alpha},\gamma_{\beta}\right\rbrace = 2 \delta_{\alpha,\beta}
  \;.
 \end{equation}
The symmetric matrix $\lambda_{\alpha,\beta}>0$ is the analogue of the coupling with the magnetic impurity in the usual Kondo problem.
This model coincides with the one studied in \cite{Altland2014}.
A related model, with real spinless fermions in the bulk, has been solved in \cite{TsvelikIsing}.

Solid-state proposals of topological Kondo devices are based on the fact that a
set of nanowires with strong Rashba coupling (InAs, InSb), laid on a 
BCS superconductor (Al,Nb) and subject to a suitably tuned magnetic field,
can develop Majorana ending modes \cite{Oreg2010,Lutchyn2010,Mourik2012}.
The TK model is obtained connecting a set of $M$ effectively one-dimensional wires
to a set of nanowires supporting Majorana modes at their ends, hosted on a mesoscopic superconducting substrate
with a large charging energy, subject to an applied electrostatic potential \cite{BeriCooper2012}. 
Interactions on the wires are not essential \cite{Altland2014multi}.
The TK effect takes place at temperatures much smaller than the charging energy and the pairing parameter
of the substrate, when only the massless Majorana modes are the relevant degrees of freedom on the island
and all the processes which change the charge of the island (hence the fermion parity) are virtual.
Under these conditions \cite{BeriCooper2012,AltlandEgger,Beri2013,Zazunov14},
the effective low-energy Hamiltonian describing the junction is the one of the TK model (\ref{eq:TopologicalKondoHamiltonian}).

This Hamiltonian has received 
considerable attention in recent years, as it can provide a realization of a non-Fermi
liquid fixed point (at temperatures much below the Kondo temperature $T_K$ \cite{hewson1997kondo}), which 
can be described by an $SO(M)_2$ Wess-Zumino-Witten conformal field theory 
\cite{Boso}. 
Remarkably, such fixed point is stable against anisotropy 
in the couplings $\lambda_{\alpha,\beta}$ \cite{BeriCooper2012},
which  in principle implies that a very accurate fine tuning of these parameters is not needed
in experiments.

Conversely, it has been observed \cite{Altland2014} that the direct coupling among Majorana modes
\cite{Altland2014multi} which originates from the overlap of their wavefunctions \cite{Oreg2010,Lutchyn2010},
can split the energies of a pair of Majorana modes.
Such a coupling is modeled by adding a term
\begin{equation}
  H_h=i\sum_{\alpha\ne\beta}h_{\alpha\beta}\gamma_{\alpha}\gamma_{\beta}
\end{equation}
to the Hamiltonian (\ref{eq:TopologicalKondoHamiltonian}), 
and its effect is that of effectively removing the pair of Majorana modes
from the zero-energy sector of the spectrum and driving the system towards another fixed point,
generated by the remaining $M-2$ modes.
This implies that in any experimental setup, the parameters
$h_{\alpha,\beta}$ must be carefully controlled. 
In condensed matter realizations, they may be controlled by having
the Majorana modes sufficiently distant from one another, given that their
wavefunctions are exponentially localized. A feature of 
the implementation proposed in this article is that 
terms involving such direct coupling do not appear without further operations 
on the system.

An exact solution of the TK Hamiltonian was proposed in \cite{Altland2014} 
for a fixed number of (noninteracting) electrons, fermion number parity,
external wires and for arbitrary coupling strength $\lambda$ (taken to be the same for all pairs of wires).
In \cite{Buccheri2015}, using the thermodynamic Bethe ansatz (TBA),
the thermodynamics of the TK Hamiltonian (\ref{eq:TopologicalKondoHamiltonian}) with an arbitrary
number of wires $M$ was investigated. 

The TBA analysis allowed to compute exactly the change in free energy due to the presence
of the coupling with Majorana modes, when the whole system is in contact with a heat bath at temperature $T$.
For an even number of wires, the free energy reads
\begin{equation}
F_{J}^{(e)}(T)=-T\sum_{j=1}^{\left\lfloor M/2 \right\rfloor}\int_{\mathbb{R}}\frac{d\omega}{2\pi}e^{i\omega/\lambda}
\frac{\cosh\frac{\omega}{2}}{\cosh\frac{(M-2)\omega}{4}}
\frac{\sinh\left(\frac{j\omega}{2}\right)}{\sinh\left(\frac{\omega}{2}\right)}
\hat L_{-,1}^{(j)}\left(\omega\right)
\end{equation}
while for an odd number of wires, the free energy is instead given by
\begin{equation}
 F_{J}^{(o)}(T)=F_{J}^{(e)}(T)+\int_\mathbb{R}\frac{d\omega }{2\pi}\frac{e^{i\omega/\lambda}\sinh\frac{(M-3)\omega}{4}}{2\cosh\frac{(M-2)\omega}{4}\sinh\frac{\omega}{2}} \hat L_{+,1}^{((M-1)/2)}(\omega)
\end{equation}
In these expressions, there appears the Fourier transform $\hat L$ of the quantity $ L_{\pm,n}^{(j)}(x)=\ln\left(1+e^{\pm \phi^{(j)}_n(x)}\right)$.
The functions $\phi^{(j)}_n(x)$ (with ``level'' index $j=1,2,\ldots, \left\lfloor M/2 \right\rfloor$ and ``length'' index $n=1,2,\ldots$)
satisfy a system of coupled nonlinear integral equations which are called TBA equations and have been written in \cite{Buccheri2015}.

In addition to that, the ground state energy shift due to the tunneling was computed to be
\begin{equation}\label{eq:GroundStateNRG}
E_{J}^{(0)}(\lambda,M)=i\ln\frac{i\Gamma\left(\frac{M+2}{4(M-2)}+\frac{i}{(M-2)\lambda}\right)\Gamma\left(\frac{3M-2}{4(M-2)}
-\frac{i}{(M-2)\lambda}\right)}{\Gamma\left(\frac{M+2}{4(M-2)}-\frac{i}{(M-2)\lambda}\right)\Gamma\left(\frac{3M-2}{4(M-2)}+\frac{i}{(M-2)\lambda}\right)}
\end{equation}
valid for both even and odd values of $M$.

The entropy introduced by the junction reduces to simple formulas both for $T\to0$ and $T\to\infty$ \cite{Altland2014,Buccheri2015}.
The zero-temperature limit is particularly noteworthy, as there appears 
a residual ground state degeneracy \cite{AffleckGroundState},
which is non-integer in the thermodynamic limit and is related to the symmetry of the fixed point. The entropy reads
\begin{equation}\label{eq:residualEntropy}
S^{(0)}_J=\ln \sqrt{\frac{M}{2}}\qquad(\mbox{even }M)\; ,\qquad\qquad S^{(0)}_J=\ln \sqrt{M}\qquad(\mbox{odd }M)\; .
\end{equation}

In order to better illustrate the peculiarity of the TK effect obtained using $Y$-junctions of 
TG gases, it is useful to discuss the junction entropy
for non-interacting fermions on the same star geometry. 
The reason is that TG and ideal Fermi gases share the same local properties 
and one may wonder whether the change in the thermodynamical properties due to the {\em junction}
may be also obtained with ideal (polarized) Fermi gases in the same geometry.
The result is that, as expected, the entropy tends to $0$ for vanishing temperature
also when the junction is present. This, as will be clear in the next section,
confirms the different junction entropy in the hard-core and ideal Fermi cases, 
as a consequence of the {\em nonlocal} nature of the couplings at the junction for the TG gases.

Finally, it was shown in \cite{Buccheri2015} that the system exhibits non-Fermi liquid
behavior at low temperatures, by computing the temperature dependence of the variation of specific heat $c_v$ due to the 
tunneling among the ends of the guides. It was found that, beside a term which is extensive in the number of particles
and linear in the temperature, the power expansion exhibits a term which is intensive and is proportional to $T^{2\frac{M-2}{M}}$,
where the characteristic power originates only from the symmetry of the low-temperature fixed point.
For the particular case $M=4$, the thermodynamic Bethe ansatz equations yield \cite{Jerez1998}
the dependence $c_v\propto -T\ln T$.

If one prepares the system in an initial state characterized by different chemical potentials on different wires,
a current $I_\alpha$ will flow from wire $\alpha$ through the junction \cite{Galpin}.
This can be characterized by the conductance tensor
$G_{\alpha,\beta}=-\frac{\partial I_\alpha}{\partial \mu_\beta}$.
At low temperatures $T\ll T_K$, in the proximity of the TK fixed point
one finds \cite{Beri2013,AltlandEgger,Altland2014multi}
%  (see \cite{BeriCooper2012} for $M=4$)
\begin{equation}
  G_{\alpha,\beta} = \frac{2Ke^2}{h}\left(\delta_{\alpha,\beta}-\frac{1}{M} \right)
  +c_{{\alpha,\beta}} T^{\frac{4}{M}}
\end{equation}
where the $c_{\alpha,\beta}$ are non universal constants and $K$ 
is the Luttinger parameter characterizing the wires \cite{Giamarchi}. 

Correlation functions of bulk operators can also be approached via bosonization.
Very close to the boundary, details of the experimental realization will be essential
and no universal prediction can be made. However, at larger distances,
but when both operators are still close to the boundary
$  x \ll \hbar v_F/T_K$,
correlation functions can be computed as in \cite{Eriksson,Bellazzini}.

\section{The \texorpdfstring{$Y$}--junction of Tonks-Girardeau gases and the topological Kondo Hamiltonian}
\label{sec:mapping}
We now prove that (\ref{eq:TopologicalKondoHamiltonian}) can be 
obtained from a model Hamiltonian for interacting bosons  in the TG limit, confined in $M$ one-dimensional waveguides 
arranged in a star geometry, i.e. a $Y$-junction. Such $Y$-junctions of low-dimensional systems 
have been studied in  a variety of physical systems. 
For Luttinger liquids crossing at a point, 
the fixed points 
\cite{ChamonPhysRevLett.91.206403,Chamon,Giuliano} and 
the transport have also been investigated \cite{Kom}. 
Lattice $Y$-junctions with non-interacting Bose gases  
were also investigated 
in \cite{Bur01,Bru04}, while in \cite{TokunoPhysRevLett.100.140402} the dynamics 
of one-dimensional Bose liquids in $Y$-junctions and the 
reflection at the center of the star was studied. 

\begin{figure}
\centering{}\includegraphics[width=0.8\textwidth]{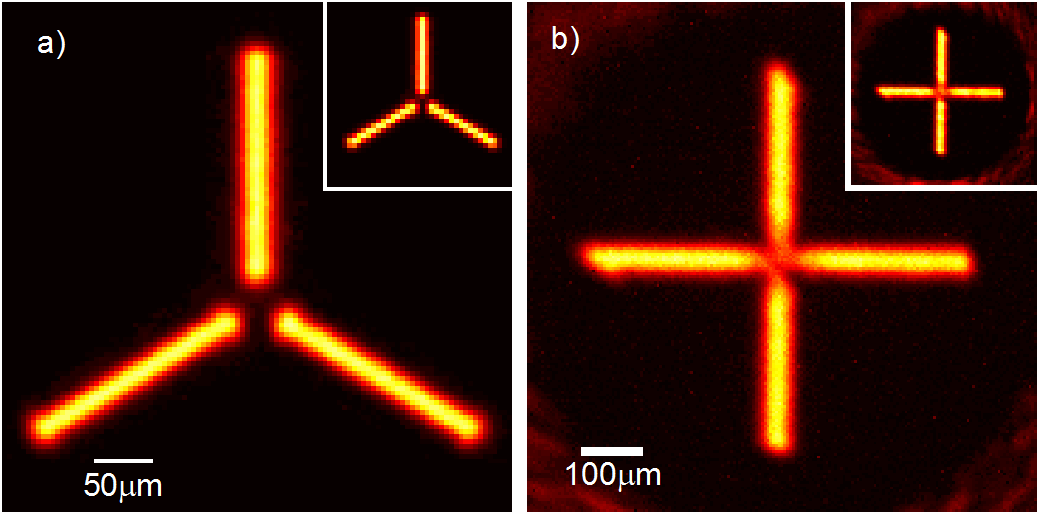}
\caption{Holographic realization of $Y$-junctions in the case of red-detuned light, with (a) three waveguides and (b) four waveguides.The large patterns are taken in the focal plane of a lens with focal length $f=$ 150 mm, while the insets are smaller patterns taken with $f=$ 50 mm. The same scale bar applies to both large and small patterns.
}
\label{fig:red}
\end{figure}

\begin{figure}
\centering{}\includegraphics[width=0.4\textwidth]{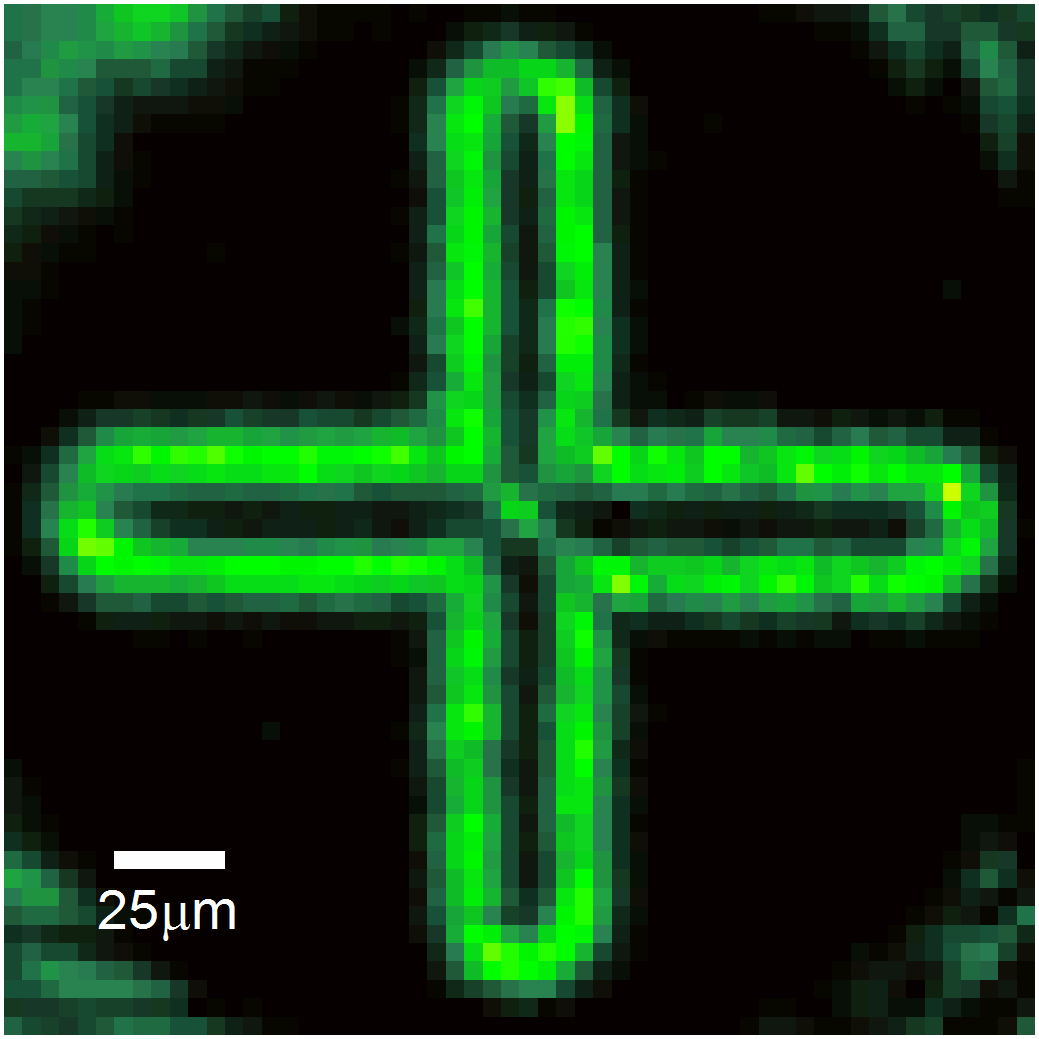} 
\caption{$Y$-junction with four waveguides in the case of blue-detuned light. The pattern is obtained with $f=$ 50 mm. The central light spot provides a barrier between the guides.
}
\label{fig:green}
\end{figure}

In figures \ref{fig:red} and \ref{fig:green}, we show optical trap geometries for ultracold bosonic atoms, which are stable and controllable $Y$-junctions of three or four waveguides connected at a point. These traps have been created using computer-generated holography, where a Spatial Light Modulator (SLM - Boulder P256, $256\times256$ pixels, 8-bit phase control) imparts a phase pattern on a 1064~nm laser beam, which is then focused by an achromatic doublet lens to create the desired intensity distribution (to achieve trapping in all directions, a light sheet can be added to provide confinement along the optical axis). The phase modulation required to realize this pattern was calculated using the Mixed-Region Amplitude Freedom algorithm \cite{Pasienski08}, and the pattern was optimized with a feedback algorithm as described in \cite{Bruce15}. The trap geometries in figure \ref{fig:red} should be implemented with red-detuned light. In this case the guides are defined by the lines of maximum intensity, and the reduced light level at the center is a potential barrier across which atoms can tunnel, thereby providing a junction term $H_J$ in the Hamiltonian. Figure \ref{fig:green} on the other hand should be implemented with blue-detuned light so that the atoms will be guided along the lines of intensity minimum, with the light spot in the center forming the potential barrier. We observe that for the four-legged patterns shown
here, the tunneling between a given pair of guides depends on whether the guides are adjacent, or on opposite sides
of the junction. However this is not a limitation, and if required it will be possible to fine-tune the tunneling, for
instance by creating multi-wavelength patterns as shown in \cite{Bowman15}.

In each waveguide $\alpha=1,\cdots,M$ the Lieb-Liniger Hamiltonian describing interacting
bosons in one-dimensional guides of length $\mathcal{L}$
reads \cite{Lieb63I,Yang69,korepin1997quantum}:
\begin{equation}
H^{(\alpha)}=\int_{0}^{\mathcal{L}}dx\left[\frac{\hbar^{2}}{2m}\partial_{x}
\Psi_\alpha^{\dagger}(x)\partial_{x}\Psi_\alpha(x)+\frac{c}{2}
\Psi_\alpha^{\dagger}(x)\Psi_\alpha^{\dagger}(x)\Psi_\alpha(x)\Psi_\alpha(x)\right]\;.
\label{eq:LiebLininger}
\end{equation}
The parameter $m$ is the mass of the bosons and $c>0$ is the repulsion strength,
as determined by the $s$-wave scattering length \cite{olshanii}.
The bosonic fields $\Psi_\alpha$ satisfy canonical commutation relations 
$\left[\Psi_\alpha(x),\Psi_\alpha^{\dagger}(y)\right]=\delta(x-y)$. 

The coupling of the Lieb-Liniger Hamiltonian, denoted by $\gamma$, is proportional 
to $c/n_{1\mathrm{D}}$ where $n_{1\mathrm{D}}\equiv\mathcal{N}/\mathcal{L}$ is the density
of bosons and $\mathcal{N}$ is the number of bosons per waveguide. More specifically we have $\gamma=mc/\hbar^2 n_{1\mathrm{D}}$. 
The limit of vanishing $c/n_{1\mathrm{D}}$ corresponds to an ideal one-dimensional Bose 
gas, while the limit of infinite $c/n_{1\mathrm{D}}$ corresponds to 
the TG gas \cite{Tonks36,Girardeau60}, which generally has the 
expectation values and thermodynamic quantities of a one-dimensional ideal Fermi gas 
\cite{korepin1997quantum,Yurovsky08,Bouchoule09,Cazalilla11}.
The experimental realization of the TG gas with cold atoms 
\cite{Paredes2009,Kinoshita2004} triggered intense 
activity in the last decade, reviewed in \cite{Yurovsky08,Bouchoule09,Cazalilla11}. 

In our experimental implementation in figure \ref{fig:green}, the width of the junction and the transverse width of the guides are close to the diffraction limit of the optical system, which is $11~\mu$m at 1064~nm. Given that the diffraction limit is proportional to wavelength, if the trap is created with light at 532~nm, all dimensions are halved. In particular, 
the full width at half maximum of the barrier becomes $d \approx 5~\mu$m. With regard to the guide parameters, 
the transverse radius of the light profile is $w_{\perp}=4~\mu$m and the length is ${\cal L}\sim 50~\mu$m. This transverse radius is sufficiently small for the gas to be in the TG regime. For instance, an optical power of the order of 
mWs creates a trap depth $D=$~500 nK and a transverse trapping frequency $\omega_{\perp}=2\sqrt{D/m}/w_{\perp}=2\pi\times$444~Hz. This is sufficient to enter the TG regime at a density $n_{1\mathrm{D}}\sim 1$ atom$/\mu$m with bosonic $^{133}$Cs atoms, given that their scattering length can be enhanced with Feshbach resonances \cite{Chu2004}. In particular $\gamma$ is given by: 
\begin{equation}\label{eq:Lieb}
 \gamma=\frac{2a_{3\mathrm{D}}}{n_{1\mathrm{D}}a_{\perp}^2}\frac{1}{1-Ca_{3\mathrm{D}}/a_{\perp}}\;,
\end{equation}
where C=1.0326, $a_{3\mathrm{D}}$ is the scattering length and $a_{\perp}=\sqrt{\hbar/(m\omega_{\perp})}$ is the harmonic oscillator length \cite{Haller09}. With $a_{3\mathrm{D}}$ = 4000 $a_0$ and our guide parameters above, we obtain $\gamma=$ 5.3, which has been shown to be within the Tonks regime both from experimental signatures \cite{Paredes2009,Kinoshita2004} and from the computation of three-body recombination rate \cite{Kormos2014}.

We now consider $M$ copies of this one-dimensional Bose gas and 
join them together by the ends of the
segments, in such a way that the bosons can tunnel from one waveguide 
to the others. The bosonic fields in different  
legs commute:
$$\left[\Psi_\alpha(x),\Psi_\beta^{\dagger}(y)\right]=
\delta_{\alpha,\beta} \,  \delta(x-y),$$ and the total Hamiltonian has the form 
$H=\sum_{\alpha=1}^{M} H^{(\alpha)}+H_J$ where the junction term 
$H_J$ describes the tunneling process among legs.

As a tool for performing computations, as well as to give a precise
meaning to the tunneling processes at the edges of the legs, 
in each leg we discretize space into a lattice of 
$L$ sites with lattice spacing $a$ (where $La=\mathcal{L}$ and the total 
number of sites $N_S$ of the star lattice is $N_S \equiv LM$). 
This discretization 
can be physically realized by superimposing optical lattices on the legs
\cite{LewensteinBook}. One can then perform a tight-binding approximation 
\cite{Jaksch98,Trombettoni01} and write the
bosonic fields as $\Psi_\alpha(x)=\sum_{\alpha, j} 
w_{\alpha, j}(x) \, b_{\alpha, j}$ where 
$b_{\alpha, j}$ is the operator destroying a particle in the site $j=1,\cdots,L$ 
of the leg $\alpha$ and $w_{\alpha, j}(x)$ 
is the appropriate Wannier wavefunction 
localized in the same site.

The resulting lattice Bose-Hubbard 
Hamiltonian on each leg then reads \cite{Jaksch98,Jaksch2005}
\begin{equation}
H_{U}^{(\alpha)} = -t\sum_{j=1}^{L-1}\left(b_{\alpha, j}^{\dagger}b_{\alpha, j+1}+
b_{\alpha, j+1}^{\dagger}b_{\alpha, j}\right)+
\frac{U}{2}\sum_{j=1}^{L}b_{\alpha, j}^{\dagger}b_{\alpha, j}^{\dagger}
b_{\alpha, j}b_{\alpha, j}\label{eq:BosonSingleChainHamiltonian}
\end{equation}
where the interaction coefficient is $U=c\int\left|w_{\alpha}(x)\right|^{4}dx$
($\alpha=1,\cdots,L$), the hopping coefficient 
is $t=-\int w_{\alpha,j} \hat{T} 
w_{\alpha,j+1} \,dx$ with $\alpha=1,\cdots,L-1$, and 
$\hat{T}=-(\hbar^2/2m)\partial^2/\partial  x^2$ is the kinetic energy operator.

The total lattice Hamiltonian, in which $M$ copies of the system are
connected to one another by a hopping term, is then written
as:
\begin{equation}\label{totalHam}
H_U=\sum_{\alpha=1}^{M}H_{U}^{(\alpha)}+H_J,
\end{equation}
where the junction term has the form 
\begin{equation}
H_J=-\lambda\sum_{1\le\alpha<\beta\le M}\left(b_{\alpha,1}^{\dagger}b_{\beta,1}+b_{\beta,1}^{\dagger}b_{\alpha,1}\right)\label{eq:BosonHamiltonian}
\end{equation}
with $\lambda$ being the hopping between the first site of a leg 
and the first sites of the others. 
Typically one has $\lambda>0$, which corresponds to an {\em antiferromagnetic} Kondo model,
as shown in the following.
Nevertheless, we observe that the sign of $t$ and
$\lambda$ could be changed by shaking the trap \cite{Eckardt}.

The total number of bosons 
in the system, $N=\mathcal{N}M$, is a conserved quantity
in the lattice model and can be tuned in experiments. 
In the canonical ensemble $N=\sum_{\alpha,j} \left\langle b_{\alpha, j}^{\dagger} b_{\alpha, j} \right\rangle$. 
The phase diagram of the bulk Hamiltonian 
(\ref{eq:BosonSingleChainHamiltonian}) in each leg 
undergoes quantum phase transitions between superfluid and Mott insulating 
phases \cite{Fisher1989}: notice that in the canonical ensemble the system 
is superfluid as soon as the filling $N/N_S$ is not integer.

We are interested in the limit $U\to\infty$, so that 
after the continuous limit is taken back again, the TG gas 
is retrieved in the bulk. It is well known that this limit brings
substantial simplifications 
in the computation: it was shown in \cite{Friedberg199352}
that, on each leg, the spectrum and the scattering matrix are equivalent 
to a system of spins in the $s=1/2$
representation. As customary, we can map the hard-core bosons to  $1/2$ spins 
by the mapping:
\begin{equation}
\label{mapping}
b_{\alpha,j}\to\sigma_{\alpha,j}^{-}\qquad b_{\alpha,j}^{\dagger}\to\sigma_{\alpha,j}^{+}\qquad2b_{\alpha,j}^{\dagger}b_{\alpha,j}-1\to\sigma_{\alpha,j}^{z}.
\end{equation}
The Hamiltonian (\ref{totalHam}) can be then written in spin form 
% $H_L=\sum_\alpha H_{\infty}^{(\alpha)} +H_J$, where:
\begin{eqnarray}
H_{\infty}^{(\alpha)} & = & -t\sum_{j=1}^{N-1}\left(\sigma_{\alpha,j}^{+}\sigma_{\alpha,j+1}^{-}+\sigma_{\alpha,j+1}^{+}\sigma_{\alpha,j}^{-}\right)\label{eq:SpinHamiltonian_a}\\
H_{J} & = & -\lambda\sum_{\alpha<\beta}^{M}\left(\sigma_{\alpha,1}^{+}\sigma_{\beta,1}^{-}+\sigma_{\beta,1}^{+}\sigma_{\alpha,1}^{-}\right)\label{eq:SpinHamiltonian} 
\end{eqnarray}
which coincides with a junction of $XX$-type spin chains \cite{Crampe2013}.

In general, only for one-dimensional systems the Jordan-Wigner transformation straightforwardly 
gives rise to a fermionic Hamiltonian which is both quadratic and local.
In order to apply the transformation also to the system at hand,
the introduction of an auxiliary spin-$1/2$ spin degree of freedom has been proposed in the case $M=3$ \cite{Crampe2013}.
This allows to write (\ref{eq:SpinHamiltonian}) as an impurity Hamiltonian, 
effectively equivalent to a four-channel Kondo model
\cite{Schlottmann1993,Tsvelick84,Tsvelick1985,Andrei1984,FabrizioGogolin}. 
For an arbitrary number of legs, the corresponding generalization 
of the Jordan-Wigner transformation is \cite{TsvelikIsing}:
\begin{equation}\label{eq:FermionsToSpins}
\sigma_{\alpha}^{-}(j)  =  \gamma_{\alpha}\prod_{l<j}e^{i\pi c_{\alpha,l}^{\dagger}c_{\alpha,l}}c_{\alpha,j}
\qquad
\sigma_{\alpha}^{z}(j)  =  2c_{\alpha,j}^{\dagger}c_{\alpha,j}-1\;,
\end{equation}
where the fermions $c_{\alpha,j}$ satisfy canonical anticommutation
relations 
\begin{equation}
\left\{ c_{\alpha,j},c_{\beta,k}^{\dagger}\right\} =\delta_{\alpha,\beta}\delta_{j,k}\qquad\left\{ c_{\alpha,j},c_{\beta,k}\right\} =0\qquad\forall\alpha,\beta\quad\forall j,k
\end{equation}
and the Klein factors $\gamma_{\alpha}$ satisfy the Clifford algebra
\begin{equation}
\left\{ \gamma_{\alpha},\gamma_{\beta}\right\} =2\delta_{\alpha,\beta}
\end{equation}

Using (\ref{eq:FermionsToSpins}) in the Hamiltonian 
(\ref{eq:SpinHamiltonian_a})-(\ref{eq:SpinHamiltonian}), one obtains
\begin{eqnarray}
H & = & -t\sum_{j=1}^{N-1}\sum_{\alpha=1}^{M}\left(c_{\alpha,j}^{\dagger}c_{\alpha,j+1}+c_{\alpha,j+1}^{\dagger}c_{\alpha,j}\right)+H_J\label{eq:TOTFER} \\
H_J & = & -\lambda\sum_{1\le\alpha<\beta\le M}\gamma_{\alpha}\gamma_{\beta}\left(c_{\alpha,1}^{\dagger}c_{\beta,1}+c_{\alpha,1}c_{\beta,1}^{\dagger}\right)\label{eq:LatticeFermionHamiltonianM}
\end{eqnarray}
In conclusion, we have mapped the Hamiltonian (\ref{eq:SpinHamiltonian}),
acting on $N_S$ spin variables, onto another one, defined in terms
of $N_S$ spinless fermionic degrees of freedom plus one Klein factor per leg. 
In other words, the hard-core boson Hamiltonian (\ref{totalHam}) in the limit $U\to\infty$
is mapped onto
the fermionic Hamiltonian (\ref{eq:TOTFER}), given by the sum of the 
$M$ non-interacting wires and the highly nontrivial junction term $H_{J}$. 
The Fermi energy of the non-interacting fermions in (each of) 
the external wires is denoted by $E_F$.

% In section \ref{sec:star2impurity}, 
We can further manipulate the coupling in the central region and conclude the mapping between 
the ferromagnetic star junction of XX spin chains and the TK model
(see also \cite{Reyes2005,Tsvelik:2014Ising,TsvelikIsing} for comparison with
Ising spin chains).
In particular, denoting by $C=\left(c_{1,1},
\ldots,
c_{M,1}
\right)^T$ the fermionic operators in the sites on the junction,
we can rewrite the interaction term for $M=3$ as an antiferromagnetic coupling
between three-flavor fermions and a localized magnetic impurity \cite{Crampe2013}
\begin{equation}
 \label{eq:M3KondoInteraction}
H_J=\lambda\sigma^{3}C^{\dagger}S^{3}C+\lambda\sigma^{2}C^{\dagger}S^{2}C+\lambda\sigma^{1}C^{\dagger}S^{1}C
\end{equation}
where $\sigma^{a}=1/(2i)\varepsilon^{abc}\gamma_{b}\gamma_{c}$, 
with $\varepsilon^{abc}$ being the completely antisymmetric tensor,
satisfy the commutation relations of the Pauli 
matrices. The matrices $S^a$ are the generators of $su(2)$ in the spin-1 representation:
\begin{equation}
S^{1}=\left(\begin{array}{ccc}
0 & 0 & 0\\
0 & 0 & -i\\
0 & i & 0
\end{array}\right)\qquad S^{2}=\left(\begin{array}{ccc}
0 & 0 & i\\
0 & 0 & 0\\
-i & 0 & 0
\end{array}\right)\qquad S^{3}=\left(\begin{array}{ccc}
0 & -i & 0\\
i & 0 & 0\\
0 & 0 & 0\\
\end{array}\right)
\end{equation}
Finally, for a general number of waveguides $M$,
after defining the impurity operators as
\begin{equation}\label{eq:so(M)impurity}
\Gamma_{\alpha,\beta}=-\frac{i}{2}\left(\gamma_{\alpha}\gamma_{\beta}-\gamma_{\beta}\gamma_{\alpha}\right)
% \qquad\qquad \sigma^{a}
\end{equation}
we obtain that the interaction can be written in the form
\begin{eqnarray}\label{eq:KondoMlattice}
H_{J} & = & \lambda\sum_{1\le\alpha<\beta\le M}\Gamma_{\alpha,\beta}C^{\dagger}T_{\alpha,\beta}C \,.
\end{eqnarray}
The operators $\Gamma_{\alpha,\beta}$ and $T_{\alpha,\beta}$ 
satisfy the $so(M)$ algebra 
\begin{equation}\label{eq:soMalgebra}
 \left[T_{\alpha,\beta},T_{\mu,\nu}\right]=-i\left(\delta_{\beta,\mu}T_{\alpha,\nu}+\delta_{\alpha,\nu}T_{\beta,\mu}-\delta_{\beta,\nu}T_{\alpha,\mu}-\delta_{\alpha,\mu}T_{\beta,\nu}\right)\;.
\end{equation}
Once again, this interaction has the form of an antiferromagnetic Kondo term. 
In extended form, the latter is just the part containing Majorana fermions of (\ref{eq:TopologicalKondoHamiltonian}).
The thermodynamic limit is taken in a standard way (see e.g. \cite{Affleck1991641}). At low temperature,
one goes back to a continuous description. The relevant physics will take place around the Fermi edges of the spectrum,
so one writes a relativistic effective Hamiltonian containing only left- and right-moving chiral fields on $M$ half-infinite
lines. The next step is to map the system onto an equivalent system in which only either left- or right-moving fields appear
and are defined between $-\infty$ and $\infty$. The junction term is located in the origin.
 The resulting effective Hamiltonian is
(\ref{eq:TopologicalKondoHamiltonian}). Note that the interaction among Majorana modes, proportional
to $h_{\alpha,\beta}$, is absent by construction.
We conclude that a $Y$-junction of TG gases provides an experimentally realizable system in which the TK effect may be observed.

We also remark that a mechanism through which an unwanted 
change of the state encoded in the Majorana modes
can occur is the loss of bosonic atoms by the trap. With our parameters, we found this possibility negligible on the time scales needed for the experiment, since the energy barrier for the atom loss is $\sim 500$~nK, much larger than the Fermi energy of the TG gases.

Another useful experimental consideration is that for the patterns shown in figures \ref{fig:red} and \ref{fig:green}, the residual intensity fluctuations at the bottom of the guides are of the order of a percent of the trap depth. This poses an upper limit to the trap depth and hence to the transverse trapping frequency that can be achieved with our off-the-shelf optics. 
However the trapping frequency could be increased by using custom-made higher resolution optics, such as in quantum gas microscopes \cite{Greiner2009,Kuhr2015,Preiss2015}, enabling us to go deeper in the TG regime. 
We also note that some arbitrary potentials have already been implemented in \cite{Preiss2015}, which suggests that holographic traps are feasible in quantum gas microscope experiments. Moreover, we could improve the signal to noise ratio in atomic images by adding an optical lattice along the axis of the optical system, therefore creating many copies of the $Y$-junction. This is similar to the technique used in the first experimental realizations of the TG limit in cold atoms \cite{Paredes2009,Kinoshita2004}, where many one-dimensional tubes are created with an optical lattice.

We finally observe that our goal was to show a method 
to obtain the TK Hamiltonian within a cold atom setup, and not to have 
a physical realization of the Kitaev chain and the associated 
Majorana modes \cite{Laflamme2014}. Indeed, in our setup, the Majorana fermions 
are not physical objects, since the observable quantities are the ones 
related to the hard-core bosons.
A related example of $Y$-junction of bosonic systems with repulsive interaction
has also appeared in \cite{YinBeri}, and connection with the topological Kondo model
has been hinted through the low-temperature behavior of the conductance.

\section{Estimates for the detection of the TK regime}
\label{sec:realization}

Next, we assess the feasibility 
of an experiment in which ultracold Cs atoms are loaded 
into the $Y$-junction presented in this paper. We provide estimates of
the tunneling coefficient $\lambda$ 
and the Kondo temperature $T_K$ given a barrier width of 
$d \sim$ 5~$\mu$m and a barrier height $V_0$ which 
can be as low as $\sim$ 15-20~nK.

An important preliminary remark is on the role of the density $n_{1\mathrm{D}}$: 
as $n_{1\mathrm{D}}$ is increased, so too is the Fermi energy $E_F$, which 
is given by $E_F=\hbar^2 \pi^2 n_{1\mathrm{D}}^2/2m$ in the one-dimensional limit (this formula is valid 
because the transverse degrees of freedom, having energies $\hbar \omega_\perp$, 
are not excited). The Kondo temperature $T_K$ is given by 
$T_K\approx E_F \exp\left\lbrace -\frac{E_F}{(M-2) \lambda} \right\rbrace$: 
for given values of $\lambda$ and $M$, $T_K$ increases with $E_F/\lambda$ up to a certain value 
of $E_F/\lambda$ (which is $\sim 1$ for $M=3$), and afterwards it
decreases with $E_F/\lambda$. At the same time, $n_{1\mathrm{D}}$
should be not too large since then $\gamma$ would decrease 
(as discussed in section \ref{sec:mapping}) and the condition 
of large $\gamma$, i.e. to have TG gases, would not be satisfied.

With these constraints in mind, we find that a density $n_{1\mathrm{D}}\sim 1$ atom/$\mu$m, giving $\gamma \approx$ 5 
and $E_F \approx$ 20~nK, leads to a favorable estimate for $T_K$.
The tunneling coefficient $\lambda$ is estimated by assuming that the 
relevant processes for the tunneling between different TG gases
only happen around $E_F$ and that the other particles only behave 
like a background Fermi sea. 
The single-particle tunneling energy $\lambda$ is the coefficient 
of the term $\propto\psi_{\beta }^{\dagger}(0)\psi_{\alpha}(0)$ representing the 
tunneling between the hard-core bosons at or near the point $0$ of 
the different bulk chains. In presence of lattices, which could be created by
superimposing lattices in each guide and merging them at the junction, 
if $\ell$ is the spacing of the lattice one can think of taking $\ell \approx d$; 
in the continuous limit it is $\ell \sim {\cal L}/{\cal N}$. It is clear that 
$\lambda$ depends on the form of the barrier, on the barrier length $d$, and 
on $V_0$ and $E_F$. Clearly, the closer $E_F$ is to $V_0$ (with $E_F<V_0$), 
the larger is $\lambda$. With $V_0-E_F$ of the order of 1-2~nK, we obtain 
$\lambda\sim$ 1-2~nK and $T_K\sim$ 10~nK. 
(We observe that tunneling through distances of $\sim 5 \mu m$ has been 
observed with Rb atoms in \cite{albiez} with an energy barrier $V_0 /h \gtrsim$ 500 Hz $\sim$ 20 nK.) 
We also remark that $V_0$
should be compared with the residual intensity fluctuations at the bottom of the waveguides which are $\sigma_V\lesssim$5~nK. Even if the two quantities are still comparable, we are reasonably within the range 
$V_0 \gg \sigma_V $ and the central barrier can be clearly resolved. 
Furthermore, with $E_F \approx$ 20~nK we also satisfy the condition 
$E_F \gg \sigma_V$, i.e. the residual intensity fluctuations 
do not significantly alter the tunneling of hard-core bosons at the Fermi 
energy scale.

The above estimates of $\lambda$ and $T_K$ are obtained 
with $d\approx$ 5~$\mu$m, so larger values may be obtained for smaller values of $d$. It is realistic 
to think that a width $d \gtrsim$ 2~$\mu$m can be obtained experimentally: as an example 
we mention the recent work \cite{Jos} in which 
a potential barrier was created with a strongly anisotropic laser beam at 532~nm focussed to a $1/e^2$ beam waist of 2~$\mu$m, corresponding to a full width at half maximum $d\approx$ 2.3~$\mu$m. Moreover,
a combined use of recently developed quantum gas microscope techniques and of holographic traps may give the possibility 
to further reduce $d$, which in turn would result in even more 
favorable estimates for the Kondo temperature.

\begin{figure}
\centering{}\includegraphics[width=0.5\textwidth]{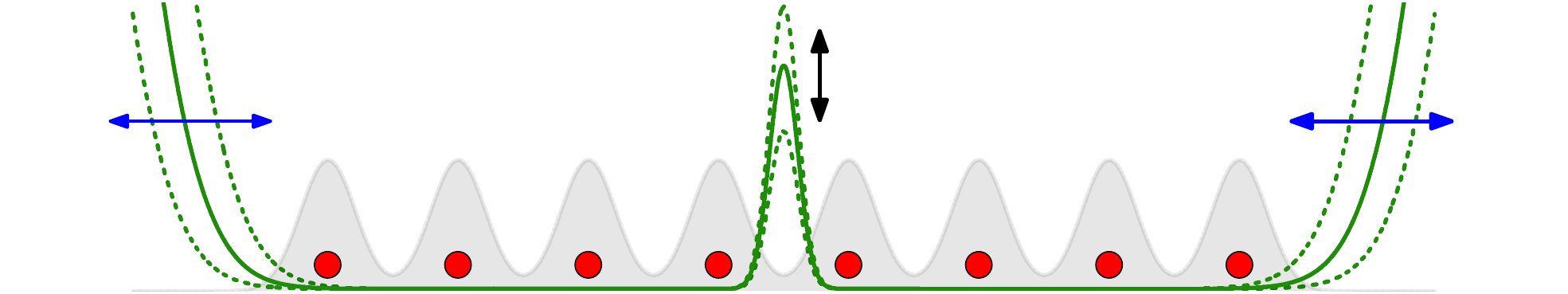} 
\caption{Pictorial representation of the parametric excitation procedure to impart energy to the TG gas, with the purpose of measuring the specific heat. The height of the barrier or the length of the guides can be modulated by changing the phase pattern on the SLM. Alternatively, the intensity of the entire pattern can be varied with an acousto-optic modulator.}
\label{fig:parametric}
\end{figure}

We finally comment on the detection of the TK regime, 
using the theoretical results reviewed in section \ref{sec:TKreview}. 
A direct way to observe the TK effect is to measure the temperature of the system
after parametric heating, from which the specific heat can be obtained. 
For this purpose, energy can be transferred to the system by modulating (with a frequency $\sim E_F$) 
the length of the guides and/or the height of the barrier, as shown in figure \ref{fig:parametric}. 
An important point is that measurements should be done with 
and without the impurity (i.e., $\lambda=0$ or $V_0 \to \infty$). Then, 
comparison of the two allows to isolate the contribution of the impurity to the 
specific heat. The low-temperature dependence has been computed using 
Bethe ansatz and has a characteristic non-Fermi-liquid contribution.
Another promising method is to extract from {\em in situ} measurements
the decay of the static density-density correlation function
around the center of the $Y$-junction, and to compare this decay 
with that of the correlation functions far from the center.
One could also perform a measurement of conductivity, where
a wavepacket is created in one of the guides and allowed to evolve across the trap.
Finally, the entropy exhibits a clear even-odd effect in $M$ for vanishing temperature.
Provided some signatures of this effect are still present at very low - yet finite - temperatures, 
one can detect these signatures from independent measurements of the temperature of the gas (extracted from time-of-flight images) and of its energy (possibly extracted from {\em in situ} images).
With the results of section \ref{sec:TKreview}, giving in practice the equation of state,
it is then possible to extract the entropy as a function of temperature and internal energy.

The concrete implementation  of these methods is certainly non-trivial,
however we are confident that with present-day technology 
a combined use of them will give clear evidence of the presence 
of the TK effect.

\section{Conclusions}
\label{sec:conclusions}

The physics of the topological Kondo (TK) effect is  based on 
the interaction of localized 
Majorana modes with external one-dimensional 
wires, merging in a star-like geometry.
In the original formulation, fermionic wires were coupled to a 
high-charging energy superconducting island, hosting a set of nanowires
with strong spin-orbit coupling which, in appropriate
magnetic field, developed localized Majorana edge modes.
Conversely, in the present paper, one-dimensional wires containing strongly
interacting bosons are coupled among them at one end. 
In particular, we considered a junction of Tonks-Girardeau (TG) 
gases arranged in a star geometry (forming a $Y$-junction).
We showed that this system provides a physical realization of the TK model. 

We then presented experimental results for 
$Y$-junctions using holographic optical 
traps.  We estimated that it is possible to have controllable 
and independent tunnelings of atoms between the different waveguides. Since 
the one-dimensional regime for cold bosons is routinely reached in ultracold 
atom experiments, both in optical traps and in atom-chip traps 
\cite{Yurovsky08,Bouchoule09,Cazalilla11}, and since TG gases 
have been widely studied in experiments \cite{Paredes2009,Kinoshita2004}, 
our holographic $Y$-junctions 
provide a proof-of-principle of the possibility of realizing TK devices 
in cold atoms experiments.
By proof-of-principle we mean that all the theoretical and experimental ingredients
(the model, the $Y$-junction and the one-dimensional TG gases) are available. 
Even though the completion of an 
ultracold TK device in the realm of atomtronics
 is certainly challenging, research in the near future
will clarify whether cold atoms may be used as a platform for studying  
the TK effect complementary to the solid-state realizations. 

We finally comment on the complementarity between the approach 
proposed in the present paper and the realization 
of the TK effect in solid state devices. 
Pros and cons of the present approach can be summarized this way: 
in the cold-atoms architecture, the interacting terms among Majorana modes
are absent, which is an advantage for the stability of the  TK effect, 
but at the same time the manipulation 
of information may be more difficult in view of the implementation of 
quantum information tasks. 

To be more specific, in solid state proposals, the charging energy of the island
must be the largest energy scale as single-particle 
tunneling onto the central island may mix fermion number parity sectors and spoil the TK effect \cite{Buccheri2015}.
Other difficulties arise from the necessity of controlling relatively strong magnetic fields on a mesoscopic scale.
Finally, one has tight bounds on the lifetime of the Majorana modes, arising from 
the poisoning of quasiparticles in the superconducting substrate, whenever the device is part of a circuit \cite{Higginbotham2015,Rainis2012}.
%  in realistic devices, 
% the lifetime of localized Majorana modes is necessarily finite
% %, due to the 
% . In particular, the phenomenon which 
% has been argued to originate the decay is the 
% quasiparticle poisoning, connected to the presence of an unpaired electron within the superconducting substrate.
In contrast, in the proposed experimental setup, the Majorana modes are nonlocally encoded in the bosonic 
field and therefore do not suffer any decoherence.
However, this implies that Majorana fermions cannot be directly 
manipulated, which, incidentally, is a problem common to many solid-state 
candidate devices for topological quantum computation. 
Indeed, in our proposal, the effective Hamiltonian is that of
the TK model, but the physical quantities to be measured are the 
observables related to the trapped atoms and not directly the Majorana modes. 
This motivates the very interesting quest of suitable 
schemes and algorithms for quantum information tasks on
devices constructed from the strongly interacting bosonic $Y$-junction.

\vspace{0.6cm}

\ack
We acknowledge very useful 
discussions with M. Burrello, R. Egger, S. Plugge, A. Tsvelik and P. Wiegmann. 
Useful correspondence with B. B\'eri is also acknowledged. P.S. and F.B.
acknowledge financial support from the Ministry of Science, Technology and
Innovation of Brazil.
G.B. and D.C. acknowledge support from UK EPSRC grant GR/T08272/01 and from the Leverhulme Trust research project grant RPG-2013-074.
A.T. is grateful to the International Institute of Physics of UFRN 
(Natal) for hospitality and to the support from the
Italian PRIN ``Fenomeni quantistici collettivi: dai sistemi fortemente correlati ai simulatori quantistici''.
H.B. acknowledges support from the Armenian grant 15T-1C308.

%\appendix
%\section*{Appendix}
%
%Figure \ref{fig:red} shows patterns with three and four waveguides to be realized with red-detuned light. In this case the guides are defined by the lines of maximum intensity, and the reduced light level at the center is a potential barrier across which atoms can tunnel, thereby providing the required junction term $H_J$. 
%\begin{figure}
%\centering{}\includegraphics[width=0.8\textwidth]{Figure-1.png}
%\caption{
%\textbf{Holographic realization of a $Y$-junction in the case of red-detuned light.}
%We show the realization of a holographic optical trap with (a) three waveguides and (b) four waveguides.
%The large patterns are taken in the focal plane of a doublet lens with 150 mm focal length, while the insets are smaller patterns taken with 50 mm focal length.
%The same scale bar applies to both large and small patterns. In the inset in (a) the 4.65 $\mu$m pixels of the CCD camera used to acquire the patterns are clearly visible.
%}
%\label{fig:red}
%\end{figure}

\section*{References}
\bibliographystyle{ieeetr}

\bibliography{Y_bosons}

\end{document}